\documentclass[sigconf, natbib=true]{acmart}

\AtBeginDocument{%
  }

\copyrightyear{2024}
\acmYear{2024}
\setcopyright{acmlicensed}
\acmConference[CIKM '24] {Proceedings of the 33rd ACM International Conference on Information and Knowledge Management}{October 21--25, 2024}{Boise, ID, USA.}
\acmBooktitle{Proceedings of the 33rd ACM International Conference on Information and Knowledge Management (CIKM '24), October 21--25, 2024, Boise, ID, USA}
\acmISBN{979-8-4007-0436-9/24/10}
\acmDOI{10.1145/3627673.3679558}

\usepackage{arydshln}
\usepackage{lipsum}
\usepackage{tabularray}
\usepackage{multirow}
\usepackage{booktabs, caption, array}
\usepackage{algorithm}
\usepackage{algorithmic}
\usepackage{comment}
\usepackage{xcolor}
\usepackage{subcaption}
\usepackage{graphicx}
\usepackage{subcaption}
\usepackage{enumitem}
\usepackage{framed}
\usepackage{hyperref} 
\usepackage{amsmath}
\definecolor{shadecolor}{gray}{0.9}
\usepackage{comment}
\usepackage{xspace}
\usepackage{colortbl}
\usepackage{xcolor}
\usepackage{bbding}
\newcommand{\mymodel}{LLM-CF\xspace}
\newcommand{\myllm}{RecGen-LLaMA\xspace}
\begin{document}

\title{Large Language Models Enhanced Collaborative Filtering}

\author{Zhongxiang Sun\textsuperscript{{$\star$}}}
\affiliation{
  \institution{Gaoling School of Artificial
Intelligence\\Renmin University of China}
  \city{Beijing}\country{China}
  }
\email{sunzhongxiang@ruc.edu.cn}

\author{Zihua Si\textsuperscript{{$\star$}}}
\affiliation{%
  \institution{Gaoling School of Artificial
Intelligence\\Renmin University of China}
  \city{Beijing}\country{China}
  }
\email{zihua_si@ruc.edu.cn}

\author{Xiaoxue Zang}
\affiliation{%
  \institution{Kuaishou Technology Co., Ltd.}
  \city{Beijing}\country{China}
  }
\email{zangxiaoxue@kuaishou.com}

\author{Kai Zheng}
\affiliation{%
  \institution{Kuaishou Technology Co., Ltd.}
  \city{Beijing}\country{China}
  }
\email{zhengkai@kuaishou.com}

\author{Yang Song}
\affiliation{%
  \institution{Kuaishou Technology Co., Ltd.}
  \city{Beijing}\country{China}
  }
\email{yangsong@kuaishou.com}

\author{Xiao Zhang}
\author{Jun Xu}
\authornote{Corresponding author. Work partially done at Engineering Research Center of Next-Generation Intelligent Search and Recommendation, Ministry of Education.
\\
Work done when Zhongxiang Sun and Zihua Si were interns at Kuaishou. 
\\ {$\star$}  Equal contribution.}
\affiliation{%
  \institution{Gaoling School of Artificial
Intelligence\\Renmin University of China}
  \city{Beijing}\country{China}
  }
\email{{zhangx89, junxu}@ruc.edu.cn}


\begin{abstract}

Recent advancements in Large Language Models (LLMs) have attracted considerable interest among researchers to leverage these models to \emph{enhance} Recommender Systems (RSs). Existing work predominantly utilizes LLMs to generate knowledge-rich texts or utilizes LLM-derived embeddings as features to improve RSs. 
Although the extensive world knowledge embedded in LLMs generally benefits RSs, the application can only 
take limited number of users and items as inputs, without adequately exploiting
collaborative filtering information. 
Considering its crucial role in RSs, one key challenge in enhancing RSs with LLMs lies in providing better collaborative filtering information through LLMs. In this paper, drawing inspiration from the in-context learning and chain of thought reasoning in LLMs, we propose the \textbf{L}arge \textbf{L}anguage \textbf{M}odels enhanced \textbf{C}ollaborative \textbf{F}iltering (\textbf{\mymodel}) framework, which distils the world knowledge and reasoning capabilities of LLMs into collaborative filtering. 
 We also explored a concise and efficient instruction-tuning method, which improves the recommendation capabilities of LLMs while preserving their general functionalities (e.g., not decreasing on the LLM benchmark).
Comprehensive experiments on three real-world datasets demonstrate that \mymodel significantly enhances several backbone recommendation models and consistently outperforms competitive baselines, showcasing its effectiveness in distilling the world knowledge and reasoning capabilities of LLM into collaborative filtering.

\end{abstract}




\maketitle

\section{Introduction}

Large Language Models (LLMs)~\cite{touvron2023LLaMA, OpenAI2023GPT4TR} have made rapid advancements, showcasing remarkable capabilities~\cite{wei2022emergent} in context comprehension, reasoning, generalization, and modeling world knowledge, and so on. 
With the advancement of Large Language Models (LLMs), many researchers are focusing on how to utilize LLMs in recommendation systems (RSs).
Many studies have already applied LLMs to various aspects of RSs, including ranking~\cite{Yue2023LLaMARecTR}, Click-Through Rate (CTR) prediction~\cite{Bao2023TALLRecAE,Dai2023UncoveringCC,Yin2023HeterogeneousKF}, sequential recommendation~\cite{Harte2023LeveragingLL}, rating prediction~\cite{Kang2023DoLU}, and data augmentation~\cite{Mysore2023LargeLM,Wei2023LLMRecLL}.
Considering specific methods to utilize LLMs for RSs,  current applications can be classified into two categories.
(1) \textbf{LLMs as RSs}: LLMs can be directly prompted or be fine-tuned to function as specialized RSs~\cite{sanner2023large,Bao2023TALLRecAE,Dai2023UncoveringCC,Yin2023HeterogeneousKF,Yue2023LLaMARecTR}. 
(2) \textbf{LLM-enhanced RSs}: Based on world knowledge and reasoning abilities, LLM-derived embedding vectors and LLM-generated texts can enhance RSs~\cite{xi2023towards,Harte2023LeveragingLL,Yin2023HeterogeneousKF,zheng2023generative,Cui2024DistillationME}. 

Despite their effectiveness, there are still several challenges to be addressed.
\textbf{LLMs as RSs} suffers from low efficiency due to the resource-intensive nature of LLMs, making their practical application challenging.
\textbf{LLM-enhanced RSs} inadequately exploit collaborative filtering information because the LLM can only take a limited number of users and items as inputs. How to better leverage LLMs to provide enhanced collaborative filtering information to existing RSs becomes key in LLM-enhanced RSs.



\begin{figure}[t]
    \centering
    \includegraphics[width=\linewidth]{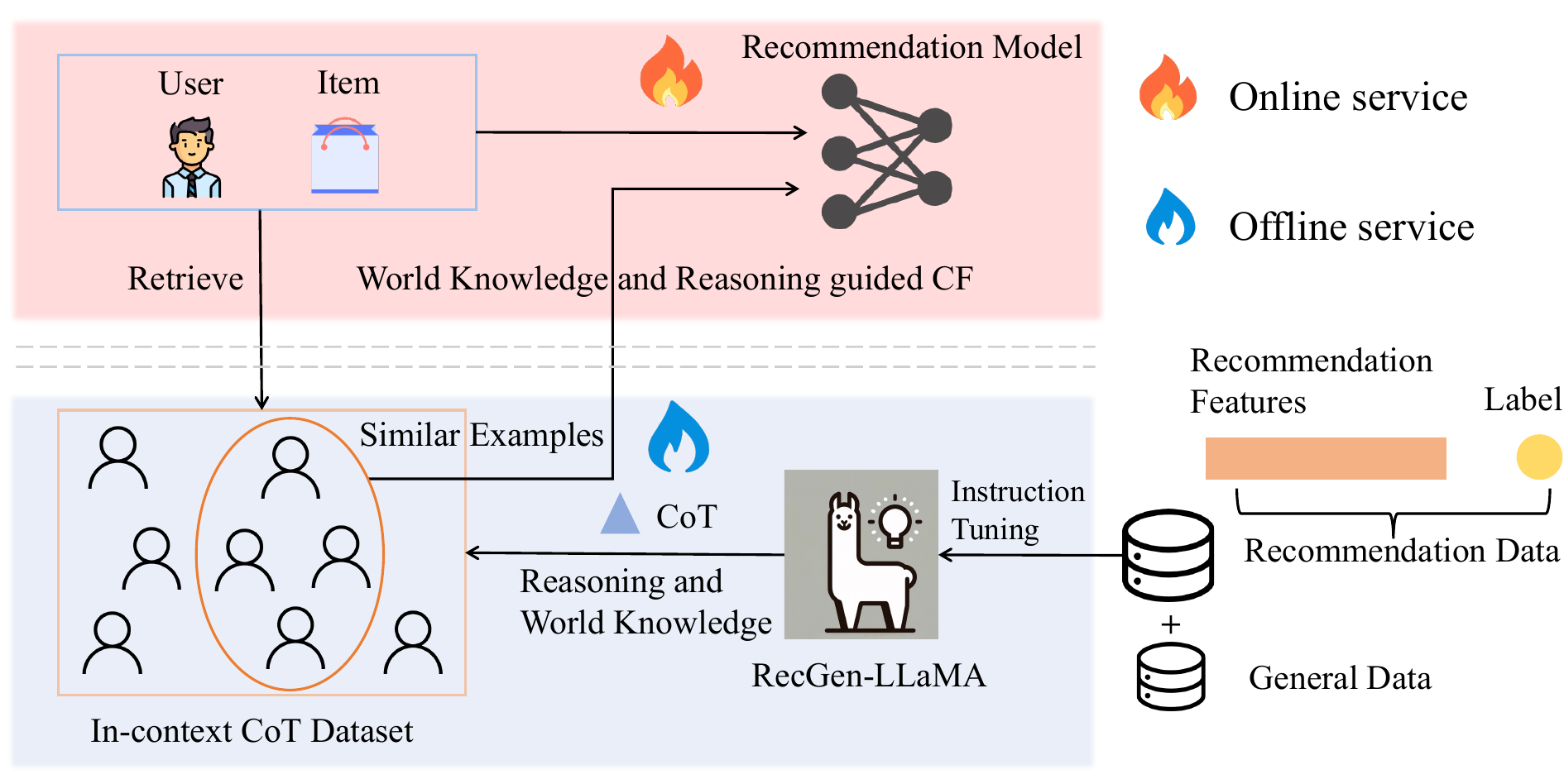}
    \caption{\mymodel integrates LLM-based world knowledge and reasoning with collaborative filtering to improve recommendation performance, using LLMs with recommender capability and decoupled latency-free offline generation.}
    \label{fig:intro}
\end{figure}

Considering the challenges in deploying LLMs as RSs due to their inherently extensive parameterization, we focus on \textbf{LLM-enhanced RSs}, which are more applicable and flexible for existing RSs. 
In order to better guide collaborative filtering to enhance existing RSs with LLMs. Inspired by Chain-of-Thought (CoT) and In-Context Learning~\cite{dong2022survey,chu2023survey} in LLMs, we propose a novel \textbf{L}arge \textbf{L}anguage \textbf{M}odels enhanced \textbf{C}ollaborative \textbf{F}iltering (\textbf{\mymodel}) Framework, which distils the world knowledge and reasoning capabilities of LLM into collaborative filtering in an in-context, chain of thought methodology. As shown in~\autoref{fig:intro}, \mymodel can be decoupled into two parts: (1) \textbf{offline service part ({\textsection~\ref{sec:offline}})}: Finetune LLM to enhance its recommendation capabilities, generate CoT reasoning with collaborative filtering information, and construct in-context CoT dataset. (2) \textbf{online service part (\textsection~\ref{sec:method}):} Retrieve the in-context CoT examples, learn the world knowledge and reasoning guided Collaborative Filtering (CF) feature, and use this feature to enhance existing RSs.

In the offline service, we perform instruction tuning on LLM to obtain CF information about users and items in the recommendation data. However, our initial findings indicate that full parameter tuning LLMs could result in substantial forgetting of their general capabilities, as discussed in \textsection~\ref{sec:sftllm}. We leveraged a simple but effective data mixing method to finetune LLaMA2~\cite{touvron2023LLaMA}, and successfully trained a model \textbf{\myllm}, which achieves an optimal balance between general and recommendation capabilities. Then, we use \myllm to generate CoT reasoning for a subset of examples in training data, forming the in-context CoT dataset.

In the online service, the retrieval module uses a query composed of the textual features of the current recommendation features to perform embedding-based retrieval on the in-context CoT dataset, forming in-context CoT examples. These retrieved examples contain similar recommendation features, as well as CoT reasoning generated by \myllm. The in-context CoT examples are concatenated with the current recommendation features and then fed into the \textbf{I}n-context \textbf{C}hain of \textbf{T}hought (\textbf{ICT}) module of \mymodel to learn world-knowledge and reasoning guided CF feature. Finally, the enhanced CF feature is fed into the backbone recommendation model for making the final prediction.


\textbf{Advantage:} \mymodel not only leverages LLMs to provide enhanced collaborative filtering information to existing RSs but also achieves exceptional deployment efficiency. (1) We teach collaborative filtering knowledge from recommendation data to LLMs, ensuring that the generated texts include a comprehensive understanding of user behaviors and item features. (2) When integrating the world knowledge and reasoning capabilities from LLMs into RSs, We design ICT modules that embody both these capabilities and collaborative filtering information. This information comes from the explicit collaborative filtering information contained in the retrieved similar user histories and the \myllm generated CoT reasoning. (3) The time cost of \mymodel is manageable, as we only require the offline serving of LLM without the need for online inference alongside the RSs.

We summarize the major contributions of this paper as follows:

(1) We represent pioneering work in using the LLMs-based world knowledge and reasoning to guide collaborative filtering features, thereby enhancing conventional recommendation models. 

(2) The proposed \mymodel is inspired by in-context learning and chain of thought reasoning in LLMs, which effectively distills the world knowledge and reasoning capabilities of LLMs into conventional recommendation models in an in-context chain of thought manner. Moreover, the \mymodel is more efficient compared to existing LLM-enhanced RSs by decoupling LLM generation from the recommendation system’s online services.

(3) We conducted extensive experiments on three public datasets. The experimental results demonstrate that the \mymodel could significantly improve the recommendation performance of conventional recommendation models in both ranking and retrieval tasks, verifying the effectiveness of the \mymodel.

\section{Related work}

\subsection{LLM as RSs}
LLMs as RSs involve directly prompting LLMs to make recommendations using natural language-based queries or adapting LLMs to serve as RSs after fine-tuning them with recommendation data. P5~\cite{p5_zhangyonggeng} transforms user interaction data into text prompts using item indices for training language models. In contrast, TALLRec~\cite{Bao2023TALLRecAE} utilizes instructional designs to outline recommendation tasks and adapts LLMs through fine-tuning to follow these guidelines, thereby producing recommendations. Further, ReLLa~\cite{lin2023rella} uses retrieved user history to fine-tune LLMs, addressing the issue of LLMs' weak capability in processing long user sequences. LLaMARec~\cite{Yue2023LLaMARecTR} initially applies small-scale recommenders to select candidates from user interaction history. Then, this history and the chosen items are fed into the LLM as text using a specially crafted prompt template. However, directly using LLMs as inference models for recommendation tasks presents challenges, such as high computational costs and slow inference times, causing challenges
in meeting the requirements for online services and deployment. 

\subsection{LLM-enhanced RSs}
LLM-enhanced RSs leverage the world knowledge and reasoning abilities of LLMs by utilizing them to generate knowledge-rich texts or employ LLM-derived embeddings as features to enhance RSs. GIRL~\cite{zheng2023generative} applies an LLM, fine-tuned with job datasets, to create job descriptions from CVs, boosting traditional job RSs. KAR~\cite{xi2023towards} uses LLMs for generating user Preference Reasoning and item Factual Knowledge, enhancing RSs through hybrid-expert adaptors. ONCE~\cite{liu2023first} explores both open and closed-source LLMs for RS enhancement; open-source LLMs as feature encoders and closed-source via prompt learning. HKF~\cite{Yin2023HeterogeneousKF} employs LLMs to merge diverse user behavior data, improving RSs by integrating these semantic features.
However, these models primarily focus on limited user-item information, neglecting collaborative filtering information, and suffer from efficiency issues due to real-time LLM processing for new interactions or items. The proposed \mymodel addresses these by distilling LLMs' knowledge and reasoning with collaborative filtering into existing RSs, separating LLM generation from online services to achieve efficient LLM-enhanced RSs.

\begin{figure}[t]
    \centering
    \includegraphics[width=0.99\linewidth]{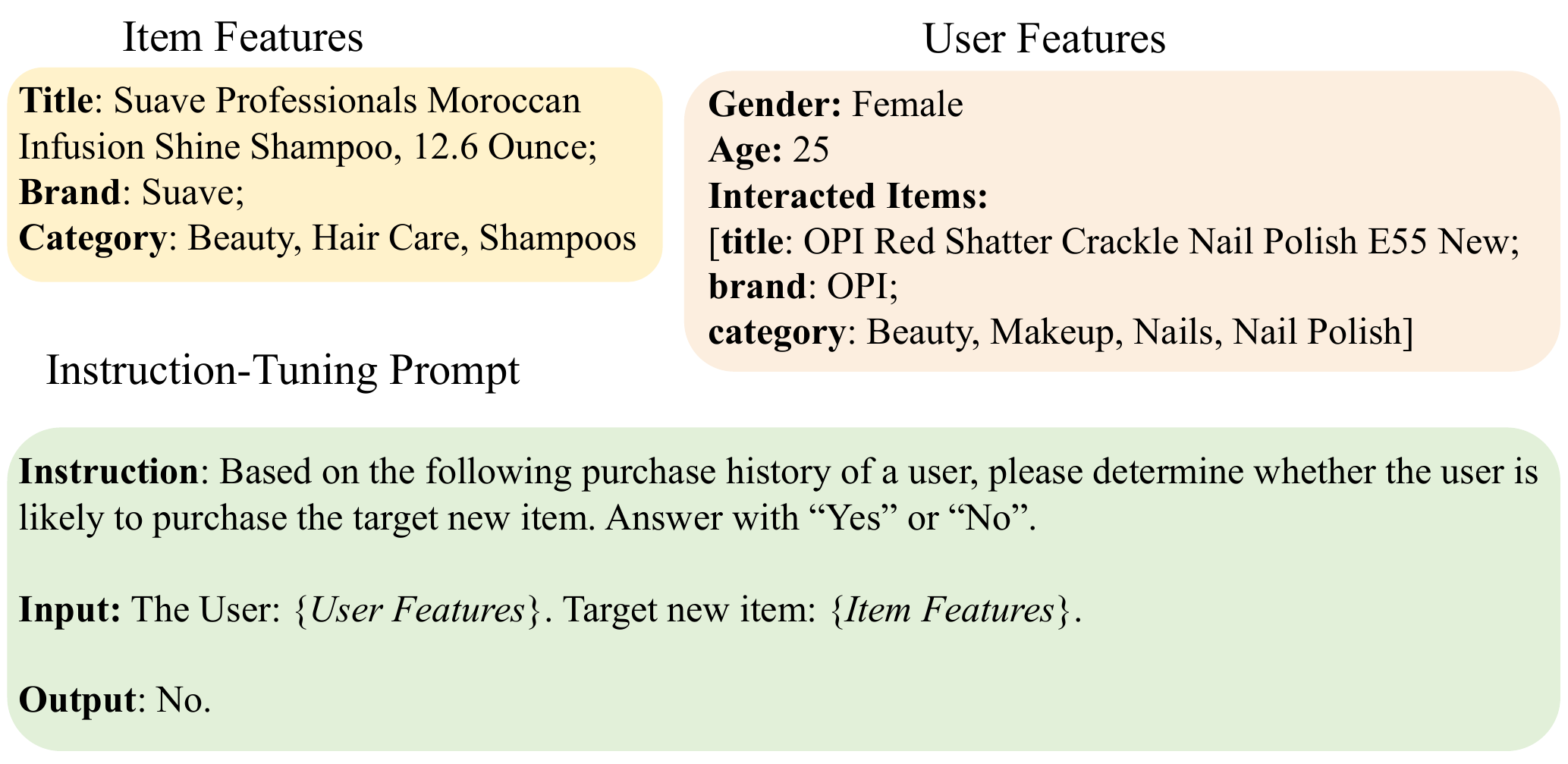}
    \caption{Example of the textual format recommendation features and Instruction-Tuning prompt. }
    \label{fig:data}
\end{figure}

\begin{figure*}[t!]
    \centering
    \includegraphics[width=\textwidth]{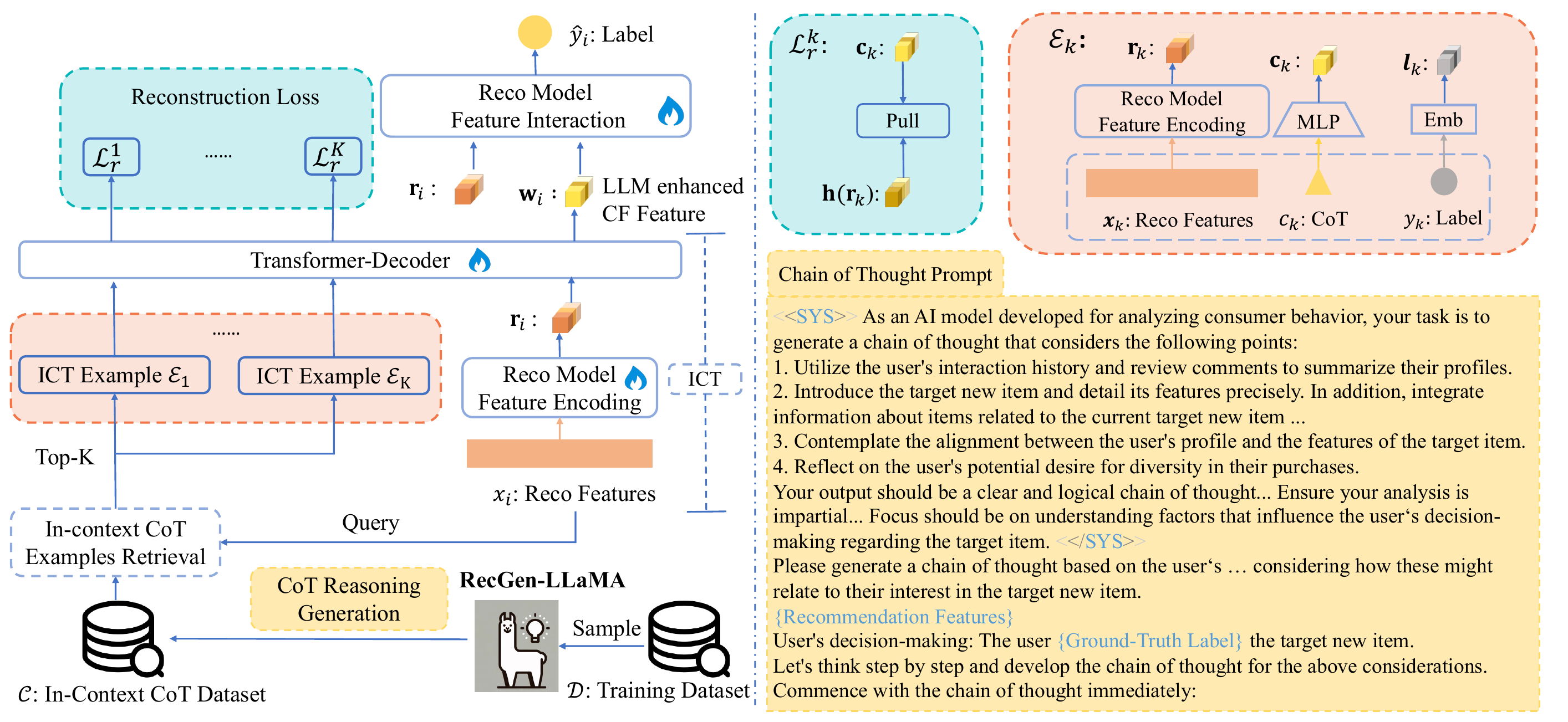}
    \caption{The overall architecture of the proposed model \mymodel. (1) CoT Reasoning Generation: Utilize Chain of Thought Prompt to generate CoT reasoning based on \myllm. (2) In-context CoT Examples Retrieval: Identify top-K similar historical recommendation data with current recommendation data to form in-context CoT examples. (3) In-context Chain of Thought (ICT) Module: Employ transformer decoder layers to learn world-knowledge and reasoning guided CF feature.}
    \label{fig:Model_Graph}
\end{figure*}

\section{PRELIMINARIES}


In this section, we define the recommendation task as a binary classification problem using multi-field categorical data. The dataset $\mathcal{D} = \{ (\mathbf{x}_n, y_n)\}_{n=1}^{N}$ comprises pairs of recommendation features $\mathbf{x}_i$ and binary labels $y_i$. Recommendation features include user features (e.g., User ID, gender, age) and item features (e.g., item ID, brand). The label $y_i$ indicates a click (1) or no click (0). The goal is to learn a function $f(\cdot)$ with parameters $\theta$ to predict click probabilities, $\hat{y}_i = f(\mathbf{x}_{i}; \theta)$, for each $\mathbf{x}_i$.

To meet the requirements of large language models, we follow the instruction prompt used in \cite{Bao2023TALLRecAE,lin2023rella}, as shown in~\autoref{fig:data}, which involves extracting textual format recommendation features $\textbf{x}_{i}^{t}$ from recommendation features $\textbf{x}_{i}$ and organizing it into a "Task Instruction". This instruction guides the LLM to determine whether a user is likely to be interested in the target item based on the user's historical interactions and the other user features. The LLM generates a binary response of "Yes" or "No", with "Yes" indicating a positive interaction (click) and "No" indicating a negative interaction (no click).

To study the world knowledge and reasoning guided collaborative filtering feature based on LLMs, we propose \mymodel (\textbf{L}arge \textbf{L}anguage \textbf{M}odels enhanced \textbf{C}ollaborative \textbf{F}iltering). The overview of \mymodel is illustrated in~\autoref{fig:Model_Graph}.  The proposed \mymodel can be decoupled into the \textbf{offline service part}~(\textbf{\textsection~\ref{sec:offline}}) and the \textbf{online service part}~(\textbf{\textsection~\ref{sec:method}}).

\section{Offline Service of \mymodel}
\label{sec:offline}
In this section, we introduce the offline service of \mymodel in detail.

\subsection{Overview}

The offline service part of \mymodel includes the following process:

 \textbf{Training of \myllm:} 
 LLMs equipped with recommendation capability can better comprehend the collaborative filtering information within recommendation data, thereby generating improved textual to enhance conventional recommendation models. However, our experiments reveal that directly following the previous work~\cite{Bao2023TALLRecAE} by using recommendation data to fine-tune LLMs leads to catastrophic forgetting of general capabilities. This results in a significant decline in the LLM benchmark to the extent that cannot generate meaningful text.
  To solve this challenge, we conducted extensive experiments and found a concise and efficient full-parameter instruction-tuning method. This method, which integrates recommendation data with general instruction-tuning data, optimizes the balance between the model's general and recommendation abilities. By applying this method to the widely-used LLaMA2~\cite{touvron2023LLaMA}, we successfully trained \textbf{\myllm}, achieving an optimal balance between general and recommendation capabilities.

 \textbf{CoT Reasoning Generation}: For the recommendation data $(\mathbf{x_{i}}, y_{i}) \in \mathcal{D}$, we designed a zero-shot CoT prompt that decomposes user-item interaction and then reconstructs them, thereafter inducing \myllm to generate CoT reasoning $c$ based on the textual representation of recommendation features $\mathbf{x}^{t}$. These $\{c_{1}, \ldots, c_{m}, \ldots, c_{M}\}$ along with the original recommendation data form the In-Context CoT Dataset $\mathcal{C} = \{ (\mathbf{x}_m, c_{m}, y_m) \}_{m=1}^{M}$.

\subsection{\myllm}
\label{sec:sftllm}

\begin{figure}
    \centering
        \begin{subfigure}{0.95\linewidth}
        \centering
        \includegraphics[width=\textwidth]{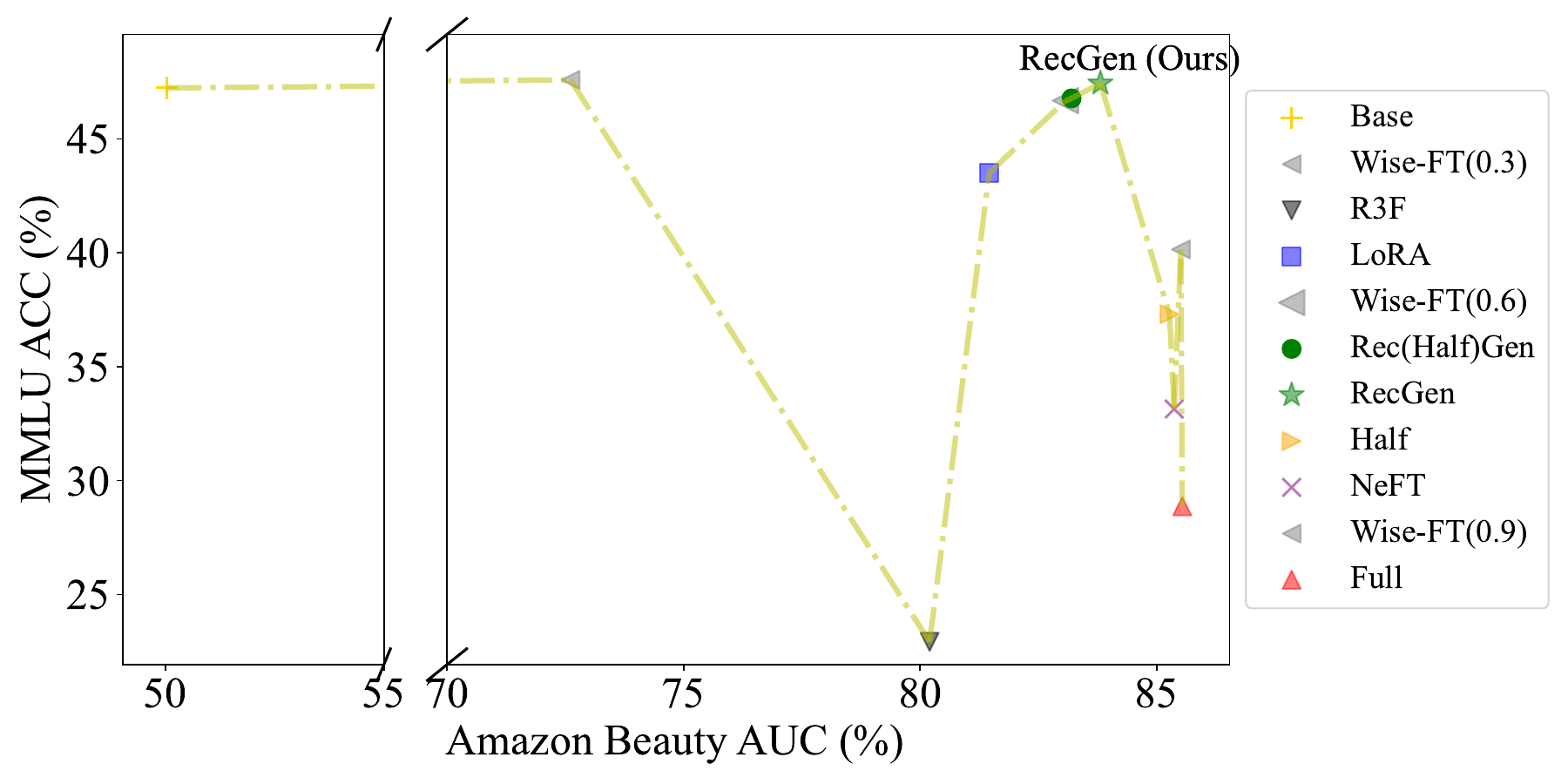}
        \label{subfig:amazon_beauty}
    \end{subfigure}

    \vspace{-8pt}

    \begin{subfigure}{0.95\linewidth}
    \centering
    \includegraphics[width=\textwidth]{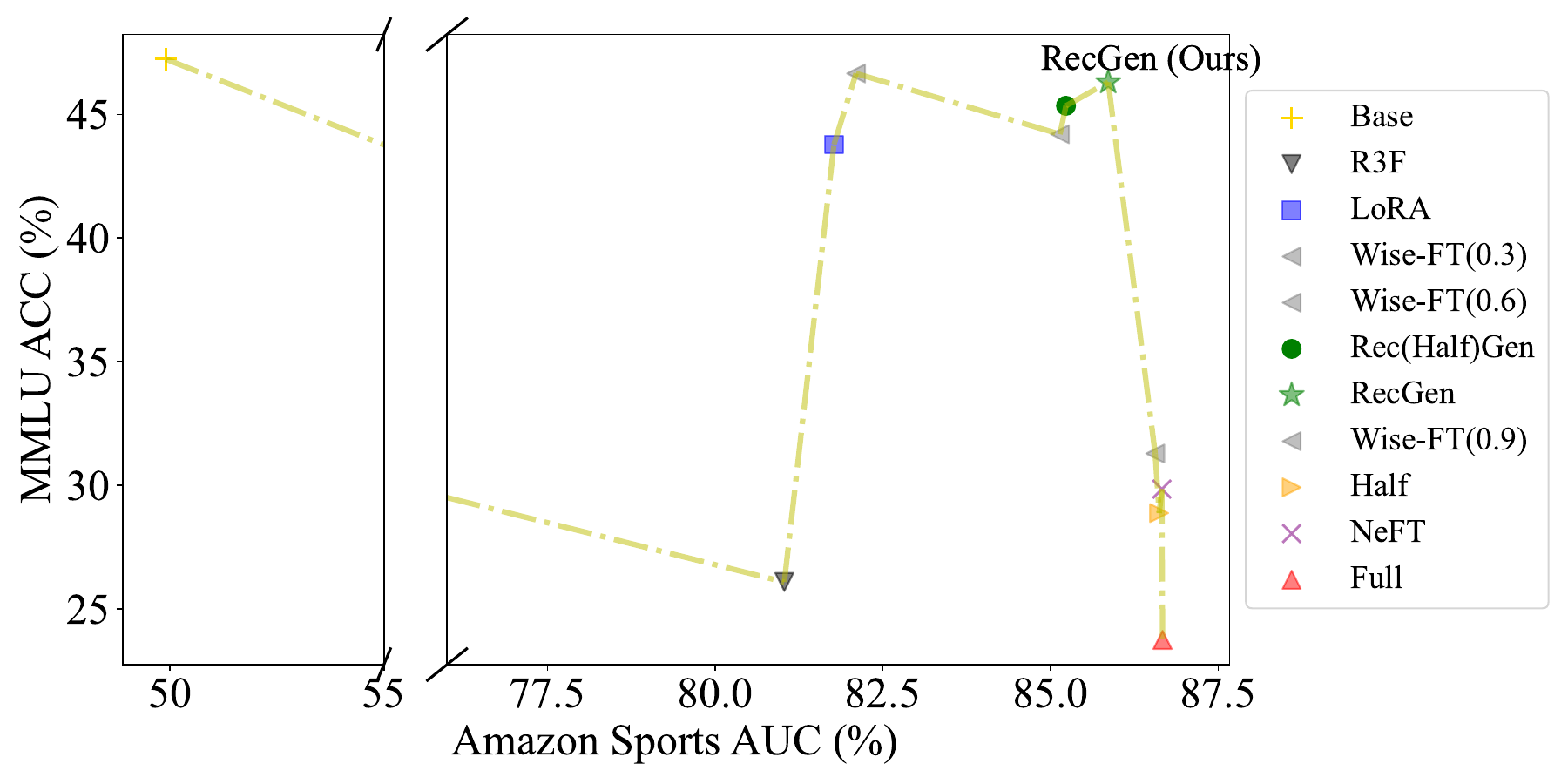}
    \label{subfig:amazon_sports}
    \end{subfigure}
    \vspace{-10pt}
    
    \caption{The recommendation and general capabilities of LLaMA2 after fine-tuning on different recommendation data using various methods.}
    \label{fig:LLM_result}
\end{figure}

 In this section, we explore how to fine-tune an LLM that balances recommendation capability with general capabilities by conducting extensive experimental analysis.

\textbf{Analysis Setup:} For the empirical study, we selected the Amazon~\cite{amazon_dataset} Beauty and Sports datasets to evaluate the recommendation capability, with detailed statistical data provided in~\textsection~\ref{sec:data}. For evaluating the general capability of LLMs, we employed the MMLU benchmark~\cite{hendrycks2020measuring}, which is widely utilized in evaluating LLM general capabilities~\cite{touvron2023LLaMA,team2023gemini,OpenAI2023GPT4TR}. The evaluation metrics include the AUC (Area Under the Curve) for recommendation capability and ACC (Accuracy) for general capability. Regarding the LLM, we chose the currently wildly used LLaMA2-7B-chat~\cite{touvron2023LLaMA} as our base model.

We systematically explore diverse methodologies to optimize LLaMA's recommendation performance while retaining its general capabilities. Our investigation encompasses Continual Learning, Robust Representation Fine-tuning, and Parameter-efficient Fine-tuning approaches: 
     \textbf{(1) Base}: Utilize LLaMA2 directly without any fine-tuning.
     \textbf{(2) Half}: Fine-tune LLaMA2 directly using half the amount of the recommendation data.
     \textbf{(3) Full}: Fine-tune LLaMA2 directly using all the recommendation data.
     \textbf{(4) LoRA}: Given that parameter-efficient fine-tuning methods can preserve most of the model parameters and thus retain the model's general capabilities, we adopted Low-Rank Adaptation~\cite{lora} on the Query and Value matrices of LLaMA2, adding parameters with a rank of 8.
     \textbf{(5) NeFT}: Adding noise during fine-tuning can prevent overfitting or representation collapse of the pre-trained model~\cite{jain2023neftune,Yuan2022HyPeBP}. We incorporated noise in the embedding layer of the input when fine-tuning LLaMA2 with recommendation data.
    \textbf{(6) R3F}: \citet{Aghajanyan2020BetterFB} proposed a Fine-tuning method rooted in an approximation to the trust region, which ensures that the pre-trained models do not forget the original pre-trained representations when they are fine-tuned for new tasks.
    \textbf{(7) Wise-FT}: \citet{Wortsman_2022_CVPR} leveraged the weighted ensemble of the pre-trained model and the fine-tuned model to enhance various capabilities of the visual models, with the ensemble weight defined as \(W_{\text{ensemble}} = \alpha \cdot W_{\text{fine-tuned}} + (1 - \alpha) \cdot W_{\text{pre-trained}}\). We ensembled the original LLaMA2 and LLaMA2 fine-tuned with recommendation data, setting \(\alpha\) to 0.3, 0.6, and 0.9 to obtain
    \textbf{Wise-FT(0.3)}, \textbf{Wise-FT(0.6)}, and \textbf{Wise-FT(0.9)} respectively.
    \textbf{(8) RecGen}: \citet{Dong2023HowAI} found that the coding and mathematical abilities of LLMs, along with their general capabilities, can be better balanced through a mixed-data fine-tuning approach. We adopt to utilize a mix of recommendation data and high-quality general data (i.e. LIMA~\cite{zhou2023lima} and alpaca-gpt4~\cite{peng2023instruction}) for fine-tuning LLaMA2.

\textbf{Results and Findings:}
We presented the results in~\autoref{fig:LLM_result}. As shown in the figure, using the off-the-shelf LLaMA2 model (\textbf{Base}) for recommendation tasks has poor performance. However, fine-tuning LLaMA2 with recommendation data (\textbf{Full}) leads to optimal recommendation performance on both Amazon Sports and Amazon Beauty, but it diminishes LLaMA2's general capabilities. We also found that \textbf{RecGen} and \textbf{Wise-FT(0.6)} achieve a good balance between general capabilities and recommendation performance, with \textbf{RecGen} showing superior results. Based on \textbf{RecGen}, we developed \textbf{\myllm} to enhance conventional recommendation models.

\subsection{CoT Reasoning Generation}
To enable \myllm to generate CoT reasoning with world knowledge and collaborative filtering information, we design the Chain of Thought prompt \(\mathcal{P}\). As shown in~\autoref{fig:Model_Graph}, this prompt can decompose user-item interactions and then reconstruct them to analyse the relation between them, aiming to analyze user data by simulating the step-by-step reasoning process of humans. Taking product recommendation as an example: Firstly, we had \myllm systematically analyze the user's interaction history and feedback comments to establish a detailed user profile. Next, \myllm introduced the target new product and its related features in detail to better understand the target product. Finally, \myllm further analyses the alignment between the user profile and the target product features, reflecting on the user's potential needs for shopping diversity.
For the training recommendation data \((\mathbf{x_{i}}, y_{i}) \in \mathcal{D}\), this generation process can be represented as:
\begin{equation}
c_{i} = \text{\myllm} (\mathbf{x_{i}}, y_i, \mathcal{P}).
\end{equation}

Considering the resource constraints in real recommendation scenarios, we can sample \(M\) training examples from \(\mathcal{D}\) for CoT reasoning generation. The sampling process can be random sampling or select representative user interaction histories. In this paper, we adopted the uniform random sampling, resulting in \(\{c_{1}, ..., c_{m}, ... , c_{M}\}\) combined with the original recommendation examples to form the In-context CoT dataset \(\mathcal{C} = \left\{ (\mathbf{x}_m, c_{m}, y_m) \right\}_{m=1}^{M}\).

\subsection{Efficiency Analysis of Offline Service}\label{sec:train_llm_eff} 

When deployed, \mymodel shows high efficiency compared to previous LLM-enhanced RSs work. It avoids real-time generation by LLMs, requiring only periodic updates on the history dataset and successfully decoupling LLM generation from the recommendation system’s online services. Next, we will analyze the time efficiency from the Training and Generation of \myllm.

\begin{table}[t]
\centering
\caption{
Comparing \mymodel with KAR in terms of the efficiency of using LLMs to generate data for dynamic scenarios. 
\textcolor{teal}{\CheckmarkBold} indicates that it does not require the real-time generation of new data; \textcolor{purple}{\XSolidBrush} indicates the opposite.
}
\begin{tabular}{ccc}
\toprule
\textbf{Scenario} & \textbf{KAR} & \textbf{\mymodel} \\
\hline
New users & \textcolor{purple}{\XSolidBrush} & \textcolor{teal}{\CheckmarkBold} \\
New items & \textcolor{purple}{\XSolidBrush} & \textcolor{teal}{\CheckmarkBold} \\
New interaction (no new items/users) & \textcolor{teal}{\CheckmarkBold} & \textcolor{teal}{\CheckmarkBold} \\
\bottomrule
\end{tabular}
\label{tab:generation}
\end{table}

\textbf{Training:} Based on the experiment results in~\autoref{fig:LLM_result}, we discovered that training with half the amount of recommendation data (i.e., \textbf{Half}) as compared to using the full dataset (i.e., \textbf{Full}) achieved better general capabilities and nearly the same recommendation performance. This phenomenon was also observed in the case of \myllm trained with half the recommendation data (i.e., \textbf{Rec(Half)Gen}). This further proves that in real-world Recommender Systems scenarios, we can use a small amount of recommendation data to fine-tune LLMs to enhance recommendation capability, thereby significantly reducing the training overhead of \myllm in real-world Recommender Systems.

\textbf{Generation:}
In the generation phase, we compared \mymodel with the currently most efficient LLM-enhanced RSs approach KAR~\cite{xi2023towards}, which achieves a certain speed-up with prestore generation when user features and item features are relatively fixed. In our efficiency analysis, we have more thoroughly considered three scenarios (i.e., new interactions, new users, new items). Our model's decoupling of generation and online recommendation effectiveness is evident from~\autoref{tab:generation}. In all three scenarios, our model does not require real-time generation to meet online recommendation needs. This is particularly significant for scenarios like short-video recommendations, where many new items appear daily, further reducing the time delay in system online services.

\section{Online Service of \mymodel}
\label{sec:method}

In this section, we introduce the online service of \mymodel in detail.

\subsection{Overview}
The online service part of \mymodel includes the following components:

     \textbf{In-context CoT Examples Retrieval}: The Retrieval process involves finding the top-$K$ historical recommendation examples similar to the current recommendation data in order to provide explicit collaborative filtering information. This involves identifying $\mathcal{I}_{i}=\{(\mathbf{x}_k, c_{k}, y_k)\}_{k=1}^{K}$ from $\mathcal{C}$, which includes recommendation data similar to the current recommendation data, as well as CoT reasoning $c_{k}$ containing world-knowledge and collaborative filtering information. 

     \textbf{In-context Chain of Thought (ICT) Module}: Inspired by the success of in-context learning and chain of thought in LLMs, we use $\mathcal{I}_{i}$ as in-context CoT examples and $\mathbf{x}_{i}$ as the query. By employing a transformer decoder layer for In-context Chain of Thought learning, we learnt the world-knowledge and reasoning guided collaborative filtering feature $\mathbf{w}_{i}$, which can be used to enhance underlying recommendation models in a model-agnostic manner.

    \textbf{Training}: During the training phase, we designed a reconstruction loss for the CoT reasoning in the in-context examples, to further strengthen the world-knowledge and reasoning capabilities contained in the collaborative filtering features generated by ICT module.

\subsection{In-context CoT Examples Retrieval}
The Retrieval module is responsible for retrieving similar In-context CoT examples for the current recommendation data $(\mathbf{x}_{i}, y_{i})$. Recent studies in retrieval-augmented recommendation have shown the potential of using current recommendation features (i.e., user features and target item features) for history retrieval to enhance collaborative filtering capabilities~\cite{Pi2020SearchbasedUI,Qin2021RetrievalI,Chang2023TWINTI}. Our approach extends this concept by not only leveraging the collaborative filtering information but also incorporating the world knowledge and reasoning abilities of \myllm.

We use the text format features $\mathbf{x}_{i}^{t}$ of $\mathbf{x}_{i}$ as the query to retrieve similar In-context CoT examples from the In-context CoT dataset $\mathcal{C}$, where the key for the examples in $\mathcal{C}$ is composed of the text format features: $\mathcal{K} = [\mathbf{x}_1^{t}, \ldots, \mathbf{x}_m^{t}, \ldots, \mathbf{x}_M^{t}]$.

We employed embedding-based retrieval, including the encoding and ranking processes. To implement this process efficiently, the retrieval process is based on the Approximate Nearest Neighbor search~\cite{Muja2009FastAN}. In the Encoding phase, we used the BGE embedding~\cite{bge_embedding} as the text encoder to convert the query $\mathbf{x}_{i}^{t}$ and the candidate keys of the In-context CoT dataset $\mathcal{K}$ into embedding formats: $$\mathbf{e}(\mathbf{x}_{i}^{t}), \mathbf{e}(\mathcal{K}) = \text{encoder}(\mathbf{x}_{i}^{t}), \text{encoder}(\mathcal{K}),$$ where $\mathbf{e}(\mathcal{K})=[\mathbf{e}(\mathbf{x}_1^{t}), \ldots, \mathbf{e}(\mathbf{x}_m^{t}), \ldots, \mathbf{e}(\mathbf{x}_M^{t})]$.

In the ranking phase, the cosine similarity between the query embedding $\mathbf{e}(\mathbf{x}_{i}^{t})$ and each candidate key's embedding in $\mathbf{e}(\mathcal{K})$ is computed. Specifically, for each embedding $\mathbf{e}(\mathbf{x}_m^{t})$ in $\mathbf{e}(\mathcal{K})$, the cosine similarity with $\mathbf{e}(\mathbf{x}_{i}^{t})$ is calculated as follows:
\[\text{sim}(\mathbf{e}(\mathbf{x}_{i}^{t}), \mathbf{e}(\mathbf{x}_m^{t})) = \frac{\mathbf{e}(\mathbf{x}_{i}^{t}) \cdot \mathbf{e}(\mathbf{x}_m^{t})}{\|\mathbf{e}(\mathbf{x}_{i}^{t})\| \|\mathbf{e}(\mathbf{x}_m^{t})\|}\]
where $\cdot$ denotes the dot product and $\|\cdot\|$ denotes the norm of a vector. Subsequently, the indices of the top-$K$ most similar candidate keys are identified based on their cosine similarity scores. These indices correspond to the examples in the In-context CoT dataset $\mathcal{C}$. The In-context CoT examples associated with these indices are extracted from $\mathcal{C}$, forming: $$\mathcal{I}_{i}=\left\{ \mathcal{E}_{1}, \ldots, \mathcal{E}_{k}, \ldots, \mathcal{E}_{K} \right\}, $$ where $\mathcal{E}_{k}=(\mathbf{x}_k, c_k, y_k).$ The $\mathcal{I}_{i}$ contains the world-knowledge and reasoning enhanced explicitly collaborative filtering information of $\textbf{x}_{i}$. 

\textbf{Remark.} (1) To prevent data leakage, we ensure that $\mathcal{I}_{i}$ will definitely not contain interaction information from future time steps of $\mathbf{x}_{i}$. (2) To ensure that the proportion of positive and negative labels in $\mathcal{I}_{i}$ does not affect the prediction results of downstream recommendation tasks, we constrain the ratio of positive and negative labels in $\mathcal{I}_{i}$ to remain equal, thereby blocking shortcuts and highlighting the effectiveness of our model.

\subsection{In-context Chain of Thought Module}
Inspired by the success of in-context learning and chain of thought in LLMs, the In-context Chain of Thought (ICT) Module learns world knowledge and reasoning-guided Collaborative Filtering features through an in-context chain of thought methodology. 

The ICT module utilizes \(\mathcal{I}_{i}\) as in-context examples and \(\mathbf{x}_{i}\) as the query, forming the ICT tokens:
\begin{equation}    
\mathbf{T} = [ \mathbf{x}_1, c_{1}, y_{1}, ...., \mathbf{x}_{K}, c_{K}, y_{K}, \mathbf{x}_i].
\end{equation}

The ICT module first encodes the recommendation features (i.e., user features and target item features), CoT reasoning, and label of $\mathbf{T}$ into corresponding tokens embedding $\mathbf{E}$:
\begin{equation}    
\mathbf{E} = [ \mathbf{r}_1, \mathbf{c}_{1}, \mathbf{l}_{1}, ...., \mathbf{r}_{K}, \mathbf{c}_{K}, \mathbf{l}_{K}, \mathbf{r}_i].
\end{equation}
The high-dimensional sparse one-hot recommendation features in $\mathbf{x}$ are mapped into low-dimensional dense space via an ID Embedding layer with an embedding lookup operation, and the text features of $\mathbf{x}$ are encoded by the text encoder (the same as the In-context CoT examples Retrieval), then the underlying recommendation model's feature encoding module encodes them into the token \(\mathbf{r}\). The CoT reasoning $c$ is also mapped into low-dimensional dense space via the text encoder and MLP projection, forming the token \(\mathbf{c}\). The label $l$ is the binary one-hot vector, which can also be mapped into a low-dimensional dense space via the ID embedding layer to obtain \(\mathbf{l}\).

Considering existing research indicating a correlation between the in-context learning capability and the structural attributes of transformer decoders~\cite{dai-etal-2023-gpt,ren2023context}, we use transformer decoder layers to encode \(\mathbf{E}\). The ICT token embedding \(\mathbf{E}\) is then fed through \(L\) Transformer Decoder blocks, generating hidden representations \(\mathbf{H}\) of ICT tokens:
\begin{equation}
     \mathbf{H} = [\mathbf{h}(\mathbf{r}_{1}), \mathbf{h}(\mathbf{c}_{1}), \mathbf{h}(\mathbf{l}_{1}), \cdots, \mathbf{h}(\mathbf{r}_{i}))] = \text{Decoder}(\mathbf{E}).
\end{equation}
The world knowledge and reasoning guided collaborative filtering feature $\mathbf{w}$ is the next token of $\mathbf{T}$ (i.e., the last hidden representations of $\mathbf{H}$ ): \[\mathbf{w}=\mathbf{h}(\mathbf{r}_{i}).\]
Please note that the length of decoder input is not long in our experiments. This ensures that the ICT module does not impose much time overhead on the recommendation models. We leave details in Section~\ref{sec: efficiency}.

\subsection{Model-Agnostic Application}
The primary goal of recommendation models is to learn a function \( f(\cdot) \) characterized by parameters \( \theta \), which can skillfully predict the click probability \( P(y_i = 1 | \mathbf{x}_i) \) for each sample \( \mathbf{x}_{i} \), formalized as \( \hat{y}_i = f(\mathbf{x}_{i}; \theta) \). The world-knowledge and reasoning guided collaborative filtering feature \( \mathbf{w}_i \) can be directly utilized in enhanced underlying recommendation models:
\begin{equation}
    \hat{y}_i = f([\mathbf{x}_{i}, \mathbf{w}_{i}]; \theta) .
\end{equation} 
For ranking models, there is a greater need for interactions between recommendation features; here, [·, ·] can be a concatenation operation, followed by the use of a feature interaction module to learn deep interactions. In contrast, retrieval models emphasize efficiency more, and [·, ·] can be an addition operation.

\subsection{Model Training}

In the model training phase, for each data \((\mathbf{x}_i, y_{i}) \in \mathcal{D}\), in addition to the original loss of the underlying recommendation model \(\mathcal{L}_o^{i}\), we also designed a reconstruction loss for the CoT reasoning in the In-context CoT Examples to further strengthen the world-knowledge and reasoning capabilities contained in the collaborative filtering features \(\mathbf{w}_{i}\). The reconstruction loss is to minimize the distance between the predicted embedding and the ground-truth embedding of CoT reasoning:
\[\mathcal{L}_{r}^{i} =  \frac{1}{K}\sum_{i=1}^{K}(1-\frac{\mathbf{c}_{i} \cdot \mathbf{h}(\mathbf{r}_{i})  }{\| \mathbf{c}_{i}\| \|\mathbf{h}(\mathbf{r}_{i})\|}),\] 
where \(K\) is the number of In-context CoT Examples.

Finally, the total loss \(\mathcal{L}\) is computed as:
\begin{equation}
    \mathcal{L} = \frac{1}{N}\sum_{i=1}^{N}(\alpha\mathcal{L}_{r}^{i}+\mathcal{L}_{o}^{i}),
\end{equation}
where \(\alpha\) are hyperparameters that control the importance of the two parts of the loss, and \(N\) is the number of recommendation data.

\subsection{Efficiency Analysis of Online Service}
\label{sec: efficiency}

In the online service of \mymodel, as shown in~\autoref{fig:Model_Graph}, only the blue fire \includegraphics[height=1.2em]{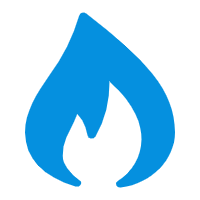} module requires online computation. 
The ICT module's inputs are short sequences that need minimal computational overhead. Experiments (\textsection~\ref{exp:ICL length}) show that the number of ICT examples only needs $K=4$, meaning the sequence length of ICT tokens $\mathbf{T}$ is only a dozen or so. Furthermore, it can utilize other acceleration techniques for transformer decoders, which are currently being widely studied~\cite{kitaev2019reformer}. The computation time of the recommendation model is essentially the same as that of underlying models, and the inputs ICT token embedding for the ICT module $\mathbf{r}$ and the final underlying model are only needed to compute once, requiring no additional computational overhead. Given the reduced necessity for maintaining an extensive In-Context CoT Dataset, the retrieval process for In-Context CoT Examples can be efficiently accelerated through the utilization of precomputed embedding vectors in conjunction with Approximate Nearest Neighbor search algorithms~\cite{Muja2009FastAN,HM-ANN}.

The analysis of online service efficiency for \mymodel, combined with the \hyperref[sec:train_llm_eff]
{offline efficiency analysis} in \textsection~\ref{sec:train_llm_eff}, demonstrates the feasibility of our framework for deployment.

\section{Experiment}

In this section, we empirically verify the effectiveness of~\mymodel by extensive experiments.
The experimental details and source codes can be found at (\textcolor{magenta}{\url{https://anonymous.4open.science/r/LLM-CF-AD78}}).

\subsection{Experimental Setup}

\subsubsection{Datasets} 
\label{sec:data}
We conducted experiments on three widely used datasets with varying domains, following existing work~\cite{p5_zhangyonggeng, CIKM2020-S3Rec, mei2023lightlm, chu2023leveraging, FMLPRec}.
Their statistics are reported in~\autoref{tab:stats}.

The Amazon\footnote{\url{http://jmcauley.ucsd.edu/data/amazon/}} review datasets~\cite{amazon_dataset} are one of the most widely used benchmarks for recommendation.
We adopt three subsets of them: \textit{Sports \& Outdoors, Beauty,} and \textit{Toys \& Games}. Following the common practices~\cite{p5_zhangyonggeng, CIKM2020-S3Rec, mei2023lightlm, FMLPRec}, we use the reviews between January 1, 2019, to December 31, 2019 and treat all the recorded reviews as positive samples. 

\begin{table}[t!]
\centering
\caption{Basic statistics of datasets.}
\begin{tabular}{lrrrr}
\toprule
Dataset &  {\textbf{Sports}} &  {\textbf{Beauty}} & {\textbf{Toys}}\\
\cmidrule{1-4}
\#Users    &  35,598   &  22,363   &  19,412  \\
\#Items    &  18,357   &  12,101  & 11,924  \\
\#Reviews  & 296,337   &  198,502   & 167,597 \\
\#Sparsity (\%) &  0.0453  & 0.0734  &  0.0724 \\
\bottomrule
\end{tabular}
\label{tab:stats}
\end{table}

\subsubsection{Baselines}

To validate the versatility of \mymodel on the ranking phase, we selected representative CTR models as backbone models. CTR prediction is commonly applied in the ranking phase of recommendation systems, focusing on modeling the probability of users clicking on items through feature interactions. The selected six \textbf{backbone} models are as follows: \textbf{DeepFM}~\cite{DeepFM}. \textbf{xDeepFM}~\cite{xDeepFM}, \textbf{AutoInt}~\cite{AutoInt}, \textbf{DCNv1}~\cite{DCNv1}, \textbf{DCNv2}~\cite{DCNv2}, and \textbf{DIN}~\cite{DIN}.

To validate the effectiveness of \mymodel, we chose representative methods that leverage LLMs to enhance recommendation models as the competitive framework. Considering that our framework initially fine-tunes LLaMA2, and LLaMA2 performs well on CTR tasks, a natural idea is to employ knowledge distillation using fine-tuned LLaMA2 to enhance CTR models. Following the previous methods~\cite{hinton2015distilling, KD4CTR}, we implemented the \textbf{KD} framework, which utilizes the logits of fine-tuned LLaMA2 as soft labels for CTR models during training. We also compared our framework with a recent work, \textbf{KAR}~\cite{xi2023towards}, which utilizes the open-world knowledge of LLMs to enhance recommendations. For a fair comparison, we used the same LLM (i.e., LLaMA2-7B-chat) as other models in \textbf{KAR}.

\subsubsection{Evaluation Metrics \& Protocal}
 Following the common practices~\cite{p5_zhangyonggeng, CIKM2020-S3Rec, mei2023lightlm, FMLPRec}, we use the \textit{leave-one-out} strategy to partition the training, validation, and test datasets. We consider all recorded interactions in the datasets as positive samples and randomly sample a negative sample for each positive sample. For the evaluation metrics, we employ widely used \textit{AUC} (Area under the ROC curve), \textit{LogLoss} (binary cross-entropy loss), and \textit{RelaImpr} (the relative improvement of AUC)~\cite{DIN} for the CTR prediction task.

\begin{table*}[ht]
    \caption{Overall performance comparisons.
    The best and the second-best performance methods are denoted in bold and underlined fonts, respectively.
    * means improvements over the second-best methods are significant (\textit{p}-value < $0.05$).
    The RelaImpr is calculated over backbone models, i.e., `None' framework. To make a fair comparison, KD, KAR, and \mymodel are all based on the 7B-parameter LLaMA2 model.
    }
    \label{exp:overall}
    \resizebox{0.93\textwidth}{!}{
         \begin{tabular}{ll|ccc|ccc|ccc}
          \toprule
          \multicolumn{1}{l}{\multirow{2}{*}{Backbone}} & \multicolumn{1}{l}{\multirow{2}{*}{Framework}} & \multicolumn{3}{c}{\textbf{Sports}}                 & \multicolumn{3}{c}{\textbf{Beauty}}  	& \multicolumn{3}{c}{\textbf{Toys}}   \\ \cmidrule(lr){3-5} \cmidrule(lr){6-8} \cmidrule(l){9-11} 
          \multicolumn{1}{c}{}                          & \multicolumn{1}{c}{}      & AUC$\uparrow$ &Logloss$\downarrow$ & RelaImpr$\uparrow$     & AUC$\uparrow$ &Logloss$\downarrow$ & RelaImpr$\uparrow$ & AUC$\uparrow$ &Logloss$\downarrow$ & RelaImpr$\uparrow$          \\          
          \midrule
          \multicolumn{1}{c}{\multirow {4}{*}{\shortstack{DeepFM}} } 
          & None & 0.7990 & 0.5471 & 0.000\% & 0.7853  & 0.5545 & 0.000\% & 0.7681 & 0.5770 & 0.000\% \\
          & KD & \underline{0.8043} & \underline{0.5404} &1.773\% &\underline{0.7959} &\underline{0.5442} & 3.715\% & \underline{0.7713} & \underline{0.5716} & 1.194\%  \\
          & KAR & 0.7991 & 0.5469 &0.033\% &0.7870 &0.5546 & 0.596\% & 0.7698 & 0.5718 & 0.634\% \\
          & \mymodel & $\textbf{0.8137}^{*}$ & $\textbf{0.5306}^{*}$ & \textbf{4.916\%}& $\textbf{0.8044}^{*}$ & $\textbf{0.5366}^{*}$ & \textbf{6.695\%} & $\textbf{0.7881}^{*}$ & $\textbf{0.5581}^{*}$ & \textbf{7.460\%} \\
          \hline
          \multicolumn{1}{c}{\multirow {4}{*}{\shortstack{xDeepFM}} } 
          & None & 0.8158 & 0.5318 & 0.000\% &0.8065 &0.5359 &0.000\% & 0.7836 & 0.5589 & 0.000\%\\
          & KD & \underline{0.8169} & 0.5298 &0.348\% &\underline{0.8104} &0.5345 &1.272\% & 0.7865 & 0.5553 & 1.023\%   \\
          & KAR & 0.8161 & \underline{0.5279} & 0.094\%&0.8101 &\underline{0.5315} & 1.175\% & \underline{0.7898} & \underline{0.5529} & 2.186\% \\
          & \mymodel & $\textbf{0.8196}^{*}$ & $\textbf{0.5248}^{*}$ &\textbf{1.203\%} & \textbf{0.8113} &\textbf{0.5311} & \textbf{1.566\%} & $\textbf{0.7947}^{*}$ & $\textbf{0.5473}^{*}$ & \textbf{3.914\%} \\
          \hline
          \multicolumn{1}{c}{\multirow {4}{*}{\shortstack{AutoInt}} } 
          & None & 0.8003 & 0.5444 &0.000\% & 0.7949 & 0.5469 & 0.000\% & 0.7630 & 0.5770 & 0.000\% \\
          & KD & 0.8012 & 0.5439 & 0.300\% &\underline{0.7961} &\underline{0.5444} & 0.407\% & 0.7635 & 0.5770 &0.190\%   \\
          & KAR & \underline{0.8039} & \textbf{0.5390} & 1.199\% &0.7939 &0.5476 & -0.339\% &\underline{0.7683} & \underline{0.5741} & 2.015\% \\
          & \mymodel & $\textbf{0.8088}^{*}$ & \underline{0.5391} & \textbf{2.831\%} &$\textbf{0.8090}^{*}$ &$\textbf{0.5321}^{*}$ & \textbf{4.781\%} & $\textbf{0.7754}^{*}$ & $\textbf{0.5685}^{*}$ & \textbf{4.714\%} \\
          \hline
          \multicolumn{1}{c}{\multirow {4}{*}{\shortstack{DCNv1}}     } 
          & None & 0.8023 & 0.5442 & 0.000\% & 0.8146 & 0.5255 & 0.000\% & 0.7621 & 0.5831 & 0.000\%\\
          & KD & \underline{0.8040} & \underline{0.5441} & 0.562\% & 0.8147 &0.5286 & 0.031\% & \underline{0.7652} & 0.5847 & 1.183\%  \\
          & KAR & 0.8024 & 0.5469 &0.033\% &\underline{0.8165}& \underline{0.5229} &0.604\% & 0.7651 & \underline{0.5821} & 1.144\% \\
          & \mymodel & $\textbf{0.8092}^{*}$ & $\textbf{0.5368}^{*}$ & \textbf{2.282\%} &$\textbf{0.8182}^{*}$ &$\textbf{0.5216}^{*}$ & \textbf{1.144\%} & $\textbf{0.7702}^{*}$ & $\textbf{0.5745}^{*}$ & \textbf{3.090\%} \\
          \hline
          \multicolumn{1}{c}{\multirow {4}{*}{\shortstack{DCNv2}} } 
          & None & 0.8110 & 0.5331 &0.000\% & 0.8028 & 0.5378 & 0.000\% & 0.7774 & 0.5650 & 0.000\% \\
          & KD & \underline{0.8112} & \underline{0.5320} &0.064\% &\textbf{0.8057} &\textbf{0.5343} & \textbf{0.958\%} & \textbf{0.7827} & \textbf{0.5609} &\textbf{1.911\%}   \\
          & KAR &  0.8087 & 0.5363 & -0.739\% &0.8003 &0.5404 & -0.825\%  & 0.7759 &	0.5662  & -0.541\% \\
          & \mymodel & $\textbf{0.8131}^{*}$ & $\textbf{0.5307}^{*}$ &\textbf{0.675\%} & \underline{0.8033} & \underline{0.5372} &0.165\% & \underline{0.7812} & \underline{0.5619} & 1.370\%\\
          \hline
          \multicolumn{1}{c}{\multirow {4}{*}{\shortstack{DIN}} } 
          & None & 0.7986 & 0.5519 & 0.000\% & 0.7861 & 0.5613 & 0.000\% & 0.7586 & 0.5885 & 0.000\% \\
          & KD & \underline{0.8023} & \underline{0.5422} & 1.239\% &\underline{0.7934} &\underline{0.5518} & 2.551\% & \underline{0.7652} & \underline{0.5847} & 2.552\%   \\
          & KAR & 0.7971 & 0.5525 & -0.502\% &0.7861&0.5604& 0.000\% & 0.7620 & 0.5874 &1.315\% \\
          & \mymodel & $\textbf{0.8089}^{*}$ & $\textbf{0.5374}^{*}$ & \textbf{3.449\%} &$\textbf{0.7967}^{*}$ &$\textbf{0.5492}^{*}$ &\textbf{3.705\%} & $\textbf{0.7783}^{*}$ & $\textbf{0.5699}^{*}$ & \textbf{7.618\%} \\
          \hline
          \bottomrule
        \end{tabular}
    }
\end{table*}

\begin{table*}[h!]
\centering
\caption{Overall performance of \mymodel on retrieval tasks. The best performance methods are denoted in bold.}
\label{exp:retrieval}
\resizebox{0.99\textwidth}{!}{
\begin{tabular}{l|cccc|cccc|cccc}
\toprule
\multicolumn{1}{l}{\multirow{2}{*}{\textbf{Models}}} & \multicolumn{4}{c}{\textbf{Beauty}} & \multicolumn{4}{c}{\textbf{Sports}} & \multicolumn{4}{c}{\textbf{Toys}} \\ 
 \cmidrule(lr){2-5} \cmidrule(lr){6-9} \cmidrule(l){10-13} 
\multicolumn{1}{c}{} & HIT@5 & HIT@10 & NDCG@5 & NDCG@10 & HIT@5 & HIT@10 & NDCG@5 & NDCG@10 & HIT@5 & HIT@10 & NDCG@5 & NDCG@10 \\ \hline
SASREC & 0.0379 & 0.0648 & 0.0206 & 0.0293 & 0.0161 & 0.0203 & 0.0088 & 0.0101 & 0.0329 & 0.0646 & 0.0143 & 0.0251 \\
+\mymodel & \textbf{0.0395} & \textbf{0.0652} & \textbf{0.0210} & 0.0293 & \textbf{0.0205} & \textbf{0.0349} & \textbf{0.0107} & \textbf{0.0154} & \textbf{0.0466} & \textbf{0.0723} & \textbf{0.0238} & \textbf{0.0321} \\ \hline
YoutubeDNN & 0.0150 & 0.0272 & 0.0089 & 0.0128 & 0.0109 & 0.0177 & 0.0069 & 0.0091 & 0.0127 & 0.0229 & 0.0077 & 0.0110 \\
+\mymodel & \textbf{0.0160} & \textbf{0.0291} & \textbf{0.0099} & \textbf{0.0139} & \textbf{0.0112} & \textbf{0.0191} & 0.0069 & \textbf{0.0094} & \textbf{0.0165} & \textbf{0.0285} & \textbf{0.0099} & \textbf{0.0138} \\ \hline
GRU4REC & 0.0279 & 0.0470 & 0.0180 & 0.0241 & 0.0165 & 0.0268 & 0.0102 & 0.0135 & \textbf{0.0257} & \textbf{0.0411} & \textbf{0.0152} & \textbf{0.0202} \\
+\mymodel & \textbf{0.0303} & \textbf{0.0504} & \textbf{0.0186} & \textbf{0.0250} & \textbf{0.0203} & \textbf{0.0317} & \textbf{0.0127} & \textbf{0.0164} & 0.0220 & 0.0380 & 0.0139 & 0.0192 \\ \hline
SRGNN & 0.0199 & 0.0341 & 0.0123 & 0.0169 & 0.0085 & 0.0146 & 0.0054 & 0.0073 & 0.0158 & 0.0284 & 0.0096 & 0.0137 \\
+\mymodel & \textbf{0.0214} & \textbf{0.0359} & \textbf{0.0128} & \textbf{0.0174} & \textbf{0.0118} & \textbf{0.0210} & \textbf{0.0072} & \textbf{0.0101} & \textbf{0.0177} & \textbf{0.0302} & \textbf{0.0104} & \textbf{0.0144} \\          
\hline
\bottomrule
\end{tabular}}
\end{table*}

 \begin{table}[h!]
    \caption{Ablation Study of \mymodel. Boldface indicates best.}
    \label{exp:ablation}
    \resizebox{0.99\linewidth}{!}{
         \begin{tabular}{ll|cc|cc|cc}
          \toprule
          \multicolumn{1}{l}{\multirow{2}{*}{Backbone}} & \multicolumn{1}{l}{\multirow{2}{*}{Framework}} & \multicolumn{2}{c}{\textbf{Sports}}                 & \multicolumn{2}{c}{\textbf{Beauty}}  & \multicolumn{2}{c}{\textbf{Toys}}   \\ \cmidrule(lr){3-4} \cmidrule(lr){5-6} \cmidrule(l){7-8} 
          \multicolumn{1}{c}{}                          & \multicolumn{1}{c}{}      & AUC$\uparrow$ &Logloss$\downarrow$     & AUC$\uparrow$ &Logloss$\downarrow$ & AUC$\uparrow$ &Logloss$\downarrow$          \\          
          \midrule
          \multicolumn{1}{c}{\multirow {6}{*}{\shortstack{DeepFM}} } 
          & None & 0.7990	& 0.5471 &0.7853&0.5545&    0.7681	&0.5770   \\
          & w/o text & \textbf{0.8137}	&0.5308&  0.8030&0.5382    &  0.7880&0.5583   \\
          & w/o CoT &  0.8009	&0.5448&0.7980	& 0.5423    &  0.7766&0.5662    \\
          & w/o \myllm &  0.8132& 0.5301 &0.7973&0.5452  &  0.7806	&0.5640  \\
          & w/o TF decoder & 0.8071&0.5378 & 0.7986&0.5411    &  0.7739	& 0.5682  \\
          & \mymodel & \textbf{0.8137}	&\textbf{0.5306}  &\textbf{0.8044}&\textbf{0.5366}   &  \textbf{0.7881}	&\textbf{0.5581}  \\
          \hline
          \multicolumn{1}{c}{\multirow {6}{*}{\shortstack{xDeepFM}} } 
          & None &  0.8158&0.5318  &  0.8065	& 0.5356  &  0.7836	&0.5589   \\
          & w/o text &  \textbf{0.8200}&0.5248  &  0.8088	& \textbf{0.5308}  &  0.7898&	0.5547   \\
          & w/o CoT &  0.8177	&0.5290  &    0.8081&	0.5349  & 0.7915&	0.5561   \\
          & w/o \myllm &  0.8191&\textbf{0.5246}  &  0.8111&	0.5311  &  0.7936&0.5490  \\
          & w/o TF decoder &  0.8192	&\textbf{0.5246}  &  0.8092&	0.5349  &  0.7912&	0.5571  \\
          & \mymodel &  0.8196&0.5248  &  \textbf{0.8113}	& 0.5311  &  \textbf{0.7947}&	\textbf{0.5473}  \\
          \hline
          \multicolumn{1}{c}{\multirow {6}{*}{\shortstack{AutoInt}} } 
          & None &  0.8003&0.5444  &  0.7949	&0.5468  &  0.7630&	0.5770   \\
          & w/o text &  0.8071&\textbf{0.5380}  &  0.8043	&0.5385  &  0.7710&	0.5725   \\
          & w/o CoT &  0.7992&0.5468  &  0.7940&	0.5490  &  0.7581&	0.5812    \\
          & w/o \myllm &  0.8041&0.5398  &  0.7993&	0.5412  &  0.7706&	0.5702  \\
          & w/o TF decoder &  0.8017&	0.5420  &  0.7942&	0.5474  &  0.7592&0.5816  \\
          & \mymodel &  \textbf{0.8088}&0.5391  &  \textbf{0.8090}	&\textbf{0.5321}  &  \textbf{0.7754}&	\textbf{0.5685}  \\
          \hline
          \multicolumn{1}{c}{\multirow {6}{*}{\shortstack{DCNv1}} } 
          & None &  0.8023&0.5442  &  0.8146&	0.5255  &  0.7621&	0.5831   \\
          & w/o text &  \textbf{0.8099}&0.5408  &  0.8155&	0.5338  &  0.7692&	0.5751   \\
          & w/o CoT &  0.8017	&0.5441  &  0.7622&	0.5802  &  0.7622&	0.5802    \\
          & w/o \myllm &  0.8036&0.5432  &  0.7891&	0.5617  &  \textbf{0.7720}&\textbf{0.5733}  \\
          & w/o TF decoder &  0.8006&0.5465  &  0.8118&0.5294  &  0.7598&	0.5858  \\
          & \mymodel &  0.8092&\textbf{0.5368}  &  \textbf{0.8182}	&\textbf{0.5216}&  0.7702&0.5745  \\
          \hline
          \multicolumn{1}{c}{\multirow {6}{*}{\shortstack{DCNv2}} } 
          & None &  0.8110&0.5331  &  0.8028	& 0.5378  &  0.7774	&0.5650   \\
          & w/o text &  0.8130&\textbf{0.5297}  &  0.7971&	0.5447  &  0.7790&	0.5628   \\
          & w/o CoT &  0.8081	&0.5365  &  0.7968&	0.5457  &  0.7753&	0.5681    \\
          & w/o \myllm &  0.8117&0.5313  &  0.8023&	0.5381  &  \textbf{0.7826}&\textbf{0.5602}  \\
          & w/o TF decoder &  0.8113&	0.5314  &  0.7978&	0.5437  &  0.7810&	0.5612  \\
          & \mymodel &  \textbf{0.8131}&0.5307  &  \textbf{0.8033}&	\textbf{0.5372}  &  0.7812&	0.5619  \\
          \hline
          \multicolumn{1}{c}{\multirow {6}{*}{\shortstack{DIN}} } 
          & None &  0.7986&0.5519  &  0.7861	& 0.5613  &  0.7586	&0.5885   \\
          & w/o text &  0.8053&0.5415  &  0.7938	&0.5509  &  0.7742&	0.5731   \\
          & w/o CoT &  0.8025	&0.5445  &  0.7907&	0.5557  &  0.7718&	0.5735    \\
          & w/o \myllm &  0.8057&0.5410  &  0.7953&	0.5571
  &  0.7783&	0.5737  \\
          & w/o TF decoder &  0.8004	&0.5484  &  0.7850&	0.5602  &  0.7666&0.5794  \\
          & \mymodel &  \textbf{0.8089}&\textbf{0.5374}  &  \textbf{0.7967}	&\textbf{0.5492}  &  \textbf{0.7783}&	\textbf{0.5699}  \\
          \hline
          \bottomrule
        \end{tabular}
    }
\end{table}

\subsubsection{Implementation Details}

As for the LLM, we utilized the widely used LLaMA2-7B-chat~\cite{touvron2023LLaMA}. 
To finetune LLaMA2, we employed the previously mentioned strategy of mixing recommendation data and general data (i.e., \textbf{RecGen} in \textsection~\ref{sec:sftllm}). We use 8 Nvidia A800 GPUs to perform full-parameter SFT on the LLM. We adopt the DeepSpeed zero-2 strategy~\cite{rajbhandari2020zero}, where each GPU has a batch size of 16, resulting in a total batch size of 128 without gradient accumulation. We used the Adam optimizer and set the initial learning rate as 1e-5 with the cosine learning rate schedule.

For the application of \mymodel to backbone models, we trained them using the Adam optimizer with a learning rate of $0.001$ and a batch size of $128$. Additionally, we adopted early-stopped training to avoid over-fitting.
The dimension of all feature embeddings is set to $32$ for all methods. The number of layers and heads in the ICT module's transformer decoder is set to  $2$. The number of In-context CoT Dataset $M$ is $\frac{1}{10}$ of the Training Dataset. The number of In-context CoT examples $K$ is set to $4$. The $\alpha$ is tuned among (0, 1] in the step of 0.1. 
Given the fact that \mymodel and competitive framework are applied over backbone models, for a fair comparison, we maintain consistency of the backbone models across all frameworks. For instance, the backbone parts of \mymodel(xDeepFM), KD(xDeepFM), and KAR(xDeepFM) share the same model architectures and basic hyper-parameters as xDeepFM.

\subsection{Experimental Results}

\autoref{exp:overall} reports the overall performance of \mymodel over 6 backbone models on three real-world datasets.
Based on the results presented in~\autoref{exp:overall},  we found 
 that \mymodel outperformed the corresponding backbone models in terms of all evaluation metrics on real-world datasets, with statistical significance. The results verified the effectiveness of the \mymodel, which can effectively integrate world-knowledge and reasoning-guided collaborative filtering features into the underlying recommendation models using the in-context chain of thought manner. 

We also observed that \mymodel achieved better results than other frameworks that leverage LLMs to enhance recommendation models (i.e., KAR and KD). Compared with \mymodel, KAR overlooks the importance of collaborative filtering information, only considering current user and item features to enhance recommendation models, and does not use an LLM fine-tuned with recommendation data, leading to a significant performance drop in real-world in-domain datasets used in our experiments. The suboptimal performance of KD is mainly because the training objective of LLMs is Cross-Entropy for all tokens, while the underlying recommendation training mainly involves Binary Cross-Entropy loss, resulting in a misalignment of optimization goals. Moreover, \mymodel is orthogonal to KD and could potentially be used in conjunction to improve recommendation performance further.
The results verify that \mymodel can better leverage LLMs to provide enhanced collaborative filtering information to underlying recommendation models. 
\begin{figure*}[t]
    \centering
    \includegraphics[width=0.90\textwidth]{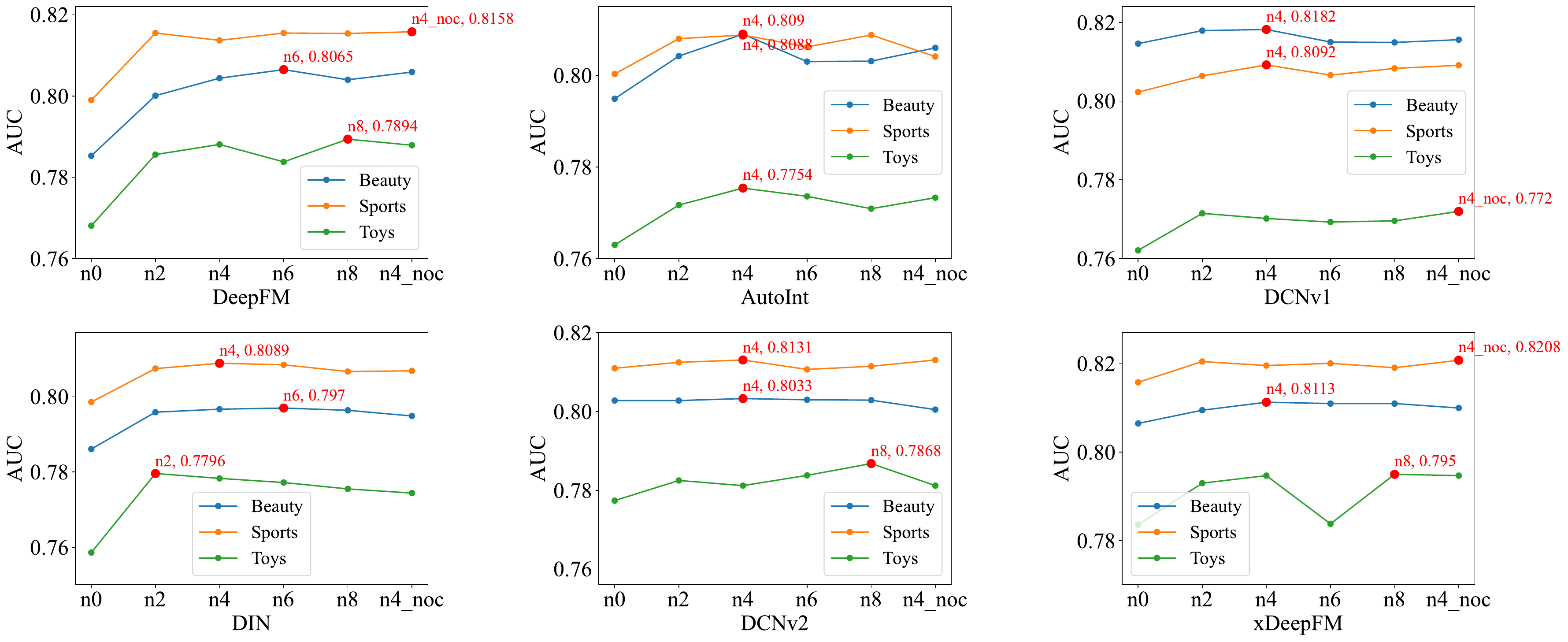}
    \caption{The recommendation performance of \mymodel on 6 backbone models across three real-world datasets as it varies with the change in In-context CoT examples length and positive/negative example constraints. The red dots indicate the best-performing setting for the current backbone model and dataset.}
    \label{fig:ICT_length}
\end{figure*}

\subsection{Ablation Study}
\label{sec:abl}

\mymodel consists of several key operations, and to understand the effects of each operation, we conducted several ablation experiments for \mymodel. They are: 
\textbf{(a) w/o text:} Remove the text features in recommendation features.
\textbf{(b) w/o CoT:} Remove the CoT reasoning along with reconstruction loss in in-context CoT examples.
\textbf{(c) w/o \myllm:} Replace the \myllm with the original LLaMA2-7B-chat to generate CoT reasoning.
\textbf{(d) w/o TF decoder:} Replace the transformer decoder layers with mean pooling in the ICT module.

Table~\ref{exp:ablation} presents the performance of various model variants over 6 backbone models on three real-world datasets.
The results reveal several insights: 
(1) After removing CoT reasoning, there is a significant decline in recommendation performance, demonstrating the effectiveness of CoT reasoning in providing world knowledge and reasoning for collaborative filtering features.
(2) Replacing \myllm with the original LLaMA2-7B-chat to generate CoT reasoning shows a downward trend in recommendation performance, highlighting the necessity of enhancing the recommendation capabilities of LLM when used to augment conventional recommendation models.
(3) After replacing the transformer decoder layers with mean pooling in the ICT module, there is a noticeable decline in recommendation performance, underscoring the importance of using transformer decoder layers for learning in-context CoT examples.
(4) Removing text features leads to a slight decline in performance, but the change is not significant, suggesting that text features have a limited role in the framework.

\subsection{Application of Retrieval Tasks}

\mymodel is versatile, applicable not only to models in the ranking stage but also to models in the retrieval stage. 
In the preceding experiments, we validated the efficacy of \mymodel in the CTR prediction task. 
In this section, we employed \mymodel over retrieval models (SASREC~\cite{kang2018self}, YoutubeDNN~\cite{Covington2016DeepNN}, GRU4REC~\cite{GRU4Rec} and SRGNN~\cite{Wu_2019}) and experimentally validated its effectiveness on four ranking metrics. 
We trained the models using the sampled softmax loss, with 128 randomly sampled negatives for each interaction.
Since these models use inner product to estimate probabilities, when applying \mymodel, we fused the vectors obtained from \mymodel with the user vector in the backbone model by element-wise addition.
To prevent data leakage, for the retrieval task, we mask the information of the target item in the current interaction in the in-context chain of thought learning.
\autoref{exp:retrieval} reports the overall performance of \mymodel over 4 backbone models on three real-world datasets. \mymodel enhances the performance of backbone retrieval models, further validating the versatility of the collaborative filtering features enhanced by world knowledge and reasoning.

\subsection{Effects of In-context CoT Examples}
\label{exp:ICL length}

Considering the length of In-context CoT (ICT) examples and the ratio of positive to negative examples has an important impact on the final prediction results. We examine the influence of the length of ICT examples and the effect of relaxing positive/negative (P/N) example ratio constraints on the performance of the \mymodel across three real-world datasets and six backbone models.

As shown in~\autoref{fig:ICT_length}, ICT lengths were varied across 0 (n0), 2 (n2), 4 (n4), 6 (n6), and 8 (n8) with a consistent P/N sample ratio. Our empirical results reveal that n4 consistently outperforms n2, while lengths of n6 and n8 show variable performance, occasionally surpassing but generally not exceeding  n4. A notable performance decline for most datasets and models at n8 suggests a diminishing return with longer ICT, positing 
 n4  as an effective balance between performance gains and computational efficiency.

Furthermore, the removal of P/N ratio constraints (denoted as \( n4\_\text{noc} \)) shown in~\autoref{fig:ICT_length} did not significantly deviate from the performance of the constrained scenarios. While some datasets showed marginal improvements without P/N constraints, others exhibited a decrease. Maintaining an equal P/N sample ratio effectively validated the superiority of \mymodel, reducing potential biases arising from disproportionate label distributions.

\section{Conclusion}
In this paper, we proposed the Large Language Model enhanced Collaborative Filtering (\mymodel) Framework, a novel approach that integrates the world knowledge and reasoning capabilities of Large Language Models into Recommender Systems. \mymodel leverages LLMs' world knowledge and reasoning, particularly through the In-context Chain of Thought module, to enhance collaborative filtering in RSs.
The key contribution of \mymodel is its effective distillation of LLM capabilities into RSs, balancing recommendation accuracy with operational efficiency. Our experiments across various datasets demonstrate that \mymodel significantly outperforms conventional recommendation models in both ranking and retrieval tasks, confirming the benefits of integrating LLMs into RSs using \mymodel.

\begin{acks}
This work was funded by the National Key R\&D Program of China (2023YFA1008704), the National Natural Science Foundation of China (No. 62377044), Beijing Key Laboratory of Big Data Management and Analysis Methods, Major Innovation \& Planning Interdisciplinary Platform for the "Double-First Class" Initiative, funds for building world-class universities (disciplines) of Renmin University of China, and PCC@RUC. Supported by Kuaishou Technology. Supported by the Outstanding Innovative Talents Cultivation Funded Programs 2024 of Renmin University of China. 
\end{acks}

\bibliographystyle{ACM-Reference-Format}
\bibliography{sample-base}

\appendix

\end{document}